\documentclass[a4paper, 9pt, twocolumn]{article}
\usepackage{geometry}                
\usepackage{amsmath} 
\usepackage{amssymb}  
\usepackage{mathtools}
\usepackage{physics}
\usepackage{authblk}
\usepackage{hyperref}

\newcommand{\R}[1]{\mathcal{R}_{#1}}       
\def\Sa{A}                         
\def\Sb{B}                         
\def\SS{X}                          
\def\sa{a}                         
\newcommand{\species}[1]{\mathrm{#1}}      
\newcommand{\Measure}{\mathbb{P}}      
\newcommand{\MeasureQ}{\mathbb{Q}} 
\newcommand{\Filtration}{\mathcal{F}}
\newcommand{\StochR}[2]{\alpha_{{#1},{#2}}}  
\newcommand{\StochP}[2]{\beta_{{#1},{#2}}}   
\newcommand{\StChange}{\nu}
\newcommand{\E}[1]{\mathbb{E}\bqty{#1}}    
\newcommand{\Path}[2]{{#1}_0^{#2}}  
\newcommand{\PMI}[3]{\mathbb{I}(\Path{#1}{#3},\Path{#2}{#3})}
\newcommand{\Mom}[2]{\Hat{#1}_{#2}}             

\def\d{\mathrm{d}}

\begin{document}

\title{
Path mutual information for a class of biochemical reaction networks
}

\author[1,2]{Lorenzo Duso}
\author[1,2,3]{Christoph Zechner}

\affil[1]{Max Planck Institute of Molecular Cell Biology and Genetics, 01307 Dresden, Germany}
\affil[2]{Center for Systems Biology Dresden, 01307 Dresden, Germany}
\affil[3]{Correspondence to: zechner@mpi-cbg.de}
\date{}
\maketitle


{\noindent \fontsize{9}{0} \selectfont \textbf{
Abstract -- Living cells encode and transmit information in the temporal dynamics of biochemical components. 
Gaining a detailed understanding of the input-output relationship in biological systems therefore requires quantitative measures that capture the interdependence between complete time trajectories of biochemical components. 
Mutual information provides such a measure but its calculation in the context of stochastic reaction networks is associated with mathematical challenges. 
Here we show how to estimate the mutual information between complete paths of two molecular species that interact with each other through biochemical reactions.
We demonstrate our approach using three simple case studies.    
}}


\section{Introduction}
The ability to continuously sense and respond to a wide spectrum of environmental signals and stresses is a hallmark of living systems. At the same time, biochemical processes inside cells are often significantly affected by random fluctuations \cite{Elowitz2002,Mcadams1997}, which stands in contrast with the remarkable robustness of biological systems. Understanding how living cells reliably transmit and process information is a fundamental problem in biology, which gained wide attention in the past. 

The combination of probabilistic methods and information theory \cite{CoverThomas} provides a rich framework to gain a quantitative understanding of signal processing in biological systems. Indeed, the use of information theoretical quantities such as the Shannon entropy and mutual information (MI) has led to insights into diverse biological systems \cite{BiophysicsBialek, Swain2014}, ranging from neural networks \cite{Bialek1998,Barbieri2004}, to biomolecular signalling and decision making systems \cite{Ziv2007,Tkacik2008, Lestas2010, Tkacik2011,Waltermann2011,Tkacik2018}.
The mutual information and related quantities such as the channel capacity have proven particularly useful in the context of biology, since they capture how information is transmitted between certain inputs (e.g., inducer molecules) and outputs (e.g., downstream targets) through cascades of biochemical reactions. 

Experimental studies have demonstrated that living cells frequently encode information in the temporal dynamics of biochemical processes \cite{Hao2012, Hansen2015}. In this case, the mutual information calculated at individual time-points would significantly underestimate the information that is transmitted across a system. A key advantage of information theoretic quantities is that they can be defined in very general terms such that they apply also to time-spanning objects. Extensions of the mutual information to trajectories have been proposed \cite{Solo2005, Guo2008, Tostevin2009, Pasha2012} as well as the closely related transfer entropy \cite{Schreiber2000, Kaiser2002, Spinney2016}. However, the analysis of path-related information measures is mathematically more demanding, especially in the context of continuous-time jump processes such as frequently encountered in biochemical systems. While the information theoretical foundation has been established for such processes \cite{Liptser2001, Solo2005, Guo2008}, the explicit calculation of the path mutual information for concrete systems remains a challenge. Previous studies have addressed this problem either numerically using particle filters \cite{Guo2008,Pasha2012} or by approximating the original jump process by a continuous Gaussian process \cite{Tostevin2009}. 

In this work we present an efficient strategy to calculate the path mutual information between two chemical species that are coupled through biochemical reactions. The method does not rely on continuous approximations of the underlying process and therefore correctly takes into account the discrete nature of biochemical networks. Our theoretical results build upon the recently proposed marginal process framework \cite{Duso2018, Bronstein2018}, which allows us to estimate the path mutual information by combining conventional stochastic simulations with the solution of a filtering problem.

The structure of the remaining paper is as follows. We review some basic definitions and concepts of stochastic chemical kinetics in Section~\ref{sec:StochasticKinetics}. In Section~\ref{sec:Goal} we formally state the problem of calculating path mutual information for a class of biochemical networks and in~\ref{sec:marginal} we show how it can be addressed using the marginal process framework.  We provide in Section~\ref{sec:RN} explicit expressions for the Radon-Nikodym derivative between the joint and marginal path measures which then allows us to state our main result in Section~\ref{sec:MainResult}. In Section~\ref{sec:previous} we briefly relate the obtained expression to previous theoretical results from the literature. Finally, we apply the method to study information transmission in three simple case studies (Section~\ref{sec:casestudies}).

\section{Theoretical Results}
\subsection{Stochastic chemical kinetics}
\label{sec:StochasticKinetics}
We consider a stochastic reaction network $\R{\SS}$ that describes the time evolution of $D$ chemical species $\species{\SS}^{(1)}, \ldots, \species{\SS}^{(D)}$ and $K$ reaction channels. The network can be defined by a set of stoichiometric equations
\begin{equation}
\sum_{i=1}^{D} \StochR{k}{i}\species{\SS}^{(i)} \xrightharpoonup{} \sum_{i=1}^{D} \StochP{k}{i}\species{\SS}^{(i)} 
\label{eq:StoichiometricEquation}
\end{equation}
for $k=1,\ldots,K$. The non-negative integer numbers $\StochR{k}{i}$ and $\StochP{k}{i}$ are respectively the reactant and product stoichiometric coefficient of species $i$ in the reaction $k$.
We introduce the stochastic process $\SS=(\SS_t)_{t\geq0}$, which takes values in $\mathbb{N}_0^D$ and is subject to the reaction dynamics in~(\ref{eq:StoichiometricEquation}). The state vector $\SS_t$ collects the discrete copy numbers of each species at time $t$ and it changes by $\StChange_k=\beta_k-\alpha_k$ when reaction $k$ occurs.
Each reaction channel $k$ is equipped with a propensity function $h_k(\SS_t)$, which sets the firing rate of this reaction. Throughout this work, we consider mass-action kinetics such that
\begin{equation}
h_k(\SS_t)=c_k \prod_{i=1}^{D} \binom{\SS_t^{(i)}}{\StochR{k}{i}},
\label{eq:MassAction}
\end{equation}
with $c_k$ as the stochastic rate constant of reaction $k$.
In this setting, $\SS$ admits a continuous-time Markov chain (CTMC), whose time evolution satisfies an integral equation of the form
\begin{equation}
	\SS_t = \SS_0 + \sum_{k=1}^K N_k\left(\int_0^t h_k(\SS_s) \mathrm{d}s\right) \StChange_k,
	\label{eq:RTC}
\end{equation}
where $N_k$ denotes a unit Poisson process counting the occurrences of reaction $k$ until time $t$ \cite{Kurtz2011}. 
We define by $\Path{\SS}{t}$ a complete path of $\SS$ collecting all information about the types and firing times of the reactions that happen between time zero and $t$. Moreover, let $\Measure^X$ denote the probability measure over the complete path $\Path{\SS}{t}$ considered on the natural filtration $\Filtration_t^X$ generated by $\SS$. Note that for a given initial condition $\SS_0$, exact realizations of $\Path{\SS}{t}$ can be simulated using the Stochastic Simulation Algorithm (SSA) \cite{gillespie1976general}.

\subsection{Information transmission between two chemical species}
\label{sec:Goal}
The goal of this work is to calculate the mutual information between complete paths of two chemical species, which we denote by $\species{\Sa}$ and $\species{\Sb}$, respectively. For simplicity, we restrict ourselves to the scenario where the network $\R{\SS}$ comprises only $\species{\Sa}$ and $\species{\Sb}$ and no additional species. The network evolves through an arbitrary number of reaction channels, whereas only reactions involving both $\species{\Sa}$ and $\species{\Sb}$ (as reactant or product) will lead to an exchange of information among the two. We refer to these reactions as \textit{coupling reactions}. Here we consider coupling reactions that modify either $\species{\Sa}$ and $\species{\Sb}$ but not both at the same time. This involves reactions of catalytic or annihilating form such as $\species{\Sa}\rightarrow \species{\Sa}+\species{\Sb}$ or $\species{\Sa}+\species{\Sb}\rightarrow \species{\Sa}$, whereas conversions such as $\species{\Sa} \rightarrow \species{\Sb}$ are not considered. As a consequence, the total set of reactions can be split into two disjoint sets $\R{\SS} = \{\R{\Sa}, \R{\Sb}\}$ where $\R{\Sa}$ and $\R{\Sb}$ are the reactions that exclusively modify $\species{\Sa}$ or $\species{\Sb}$, respectively. In the following, we refer to this property as \textit{smooth coupling} between $\species{\Sa}$ and $\species{\Sb}$.

Without loss of generality, we arrange the state vector such that $\SS_t=\pqty{\Sa_t,\Sb_t}$. We define the \textit{partial} paths $\Path{\Sa}{t}$ and $\Path{\Sb}{t}$ and corresponding sub-filtrations $\Filtration_t^A, \Filtration_t^B\subset \Filtration_t^X=\Filtration_t^{AB}$ generated by $\Sa_t$ and $\Sb_t$. Notice that since $\species{\Sa}$ and $\species{\Sb}$ are smoothly coupled, 
$\Filtration_t^A$ and $\Filtration_t^B$ contain information exclusively about reactions in $\R{\Sa}$ and $\R{\Sb}$, respectively. Using these definitions, our goal is to characterize information transmission between paths $\Path{\Sa}{t}$ and $\Path{\Sb}{t}$ via their path mutual information (path-MI)
\begin{equation}
    \PMI{\Sa}{\Sb}{t} = \E{\log \frac{\d\Measure^{\Sa \Sb}}{\d(\Measure^\Sa \times \Measure^\Sb)}}.
    \label{eq:PMI}
\end{equation}
In (\ref{eq:PMI}), the term inside the logarithm corresponds to the Radon-Nikodym derivative of the joint path measure $\Measure^{AB}$ with respect to the product of the \textit{marginal} path measures $\Measure^{A}$ and $\Measure^B$. The latter can be thought of as the probability laws that describe the time evolution of \textit{only} $\species{\Sa}$ or $\species{\Sb}$, respectively. We and others have previously shown how such marginal process models can be constructed for general reaction networks \cite{Zechner2014b,Duso2018}. While in those studies, it was used predominantly for the purpose of model reduction and stochastic simulation, it will now serve us to find explicit expressions of the Radon-Nikodym derivative in~(\ref{eq:PMI}). In the following section, we will show how the marginal process dynamics can be obtained for the considered two-species network.

\subsection{Marginal process dynamics}
\label{sec:marginal}
As a first step, we instantiate the dynamics from (\ref{eq:RTC}) for the considered network. This yields two coupled integral equations of the form
\begin{align}
    \Sa_t &= \Sa_0 +\sum_{k\in \R{\Sa}} N_k\left(\int_0^t \lambda_k^{AB}(s) \d s\right) \StChange_k^{\Sa} \label{eq:RTCReduced_CondA}\\
    \Sb_t &= \Sb_0 +\sum_{k\in \R{\Sb}} N_k\left(\int_0^t \lambda_k^{AB}(s) \d s\right) \StChange_k^{\Sb} ,
	\label{eq:RTCReduced_CondB}
\end{align}
where $\StChange^{\Sa}_k$ and $\StChange^{\Sb}_k$ are respectively the stoichiometric change coefficients corresponding to $\species{\Sa}$ and $\species{\Sb}$. Note that we have introduced $\lambda_k^{AB}(t)=h_k(A_t,B_t)$ to emphasize that
(\ref{eq:RTCReduced_CondA}) and~(\ref{eq:RTCReduced_CondB}) are the equations of motion for $\Sa_t$ and $\Sb_t$ relative to their (joint) filtration $\Filtration_t^{AB}$. The corresponding solutions $\Path{\SS}{t}=\{\Path{\Sa}{t}, \Path{\Sb}{t}\}$ admit a joint measure $\Measure^{AB}$. Our goal is to find two analogous equations, each being consistent with the marginal path measures $\Measure^{A}$ and $\Measure^{B}$, respectively. Technically, this can be achieved by restricting the two processes to depend only on their own history (i.e., $\Filtration_t^A$ in case of $\Sa_t$ and $\Filtration_t^B$ in case of $\Sb_t$) but not their joint history captured by $\Filtration_t^{AB}$. Informally, this can be understood as "integrating out" the dependency of one path on the other one. It can be shown \cite{Duso2018} that relative to the filtrations $\Filtration_t^{A}$ and $\Filtration_t^{B}$, the counting processes~(\ref{eq:RTCReduced_CondA}) and~(\ref{eq:RTCReduced_CondB}) evolve according to
\begin{align}
    \Sa_t &= \Sa_0 +\sum_{k\in \R{\Sa}} N_k\left(\int_0^t \lambda_k^\Sa(s) \d s\right) \StChange_k^{\Sa} \label{eq:RTCReduced_MargA}\\
    \Sb_t &= \Sb_0 +\sum_{k\in \R{\Sb}} N_k\left(\int_0^t \lambda_k^\Sb(s) \d s\right) \StChange_k^{\Sb}, \label{eq:RTCReduced_MargB}
\end{align}
where we have defined $\lambda_k^\Sa(t) = \E{h_k(\Sa_t, \Sb_t) \mid \Filtration_t^{\Sa}}$ and  $\lambda_k^\Sb(t) = \E{h_k(\Sa_t, \Sb_t) \mid \Filtration_t^{\Sb}}$. In other words, the original propensities $h_k$ are replaced by their expected value conditionally on the sub-filtrations $\Filtration_t^{\Sa}$ and $\Filtration_t^{\Sb}$. In the information theory literature, these conditional expectations are commonly referred to as \textit{causal} or \textit{filtering} estimates, since they provide a reconstruction of the original propensities $h_k$ at time $t$ from a continuous path up to time $t$ (i.e., either $\Path{\Sa}{t}$ or $\Path{\Sb}{t}$). Notice that due to the marginalization, (\ref{eq:RTCReduced_MargA}) and~(\ref{eq:RTCReduced_MargB}) are now decoupled from each other, such that their solutions are consistent with the product measure $\Measure^A\times\Measure^B$. We point out, however, that the dynamics are no longer Markovian due to the history-dependence of $\lambda_k^\Sa(t)$ and $\lambda_k^\Sb(t)$. 

While~(\ref{eq:RTCReduced_MargA}) and~(\ref{eq:RTCReduced_MargB}) provide the marginal descriptions we seek for, the calculation of the marginalized propensities $\lambda_k^\Sa(t)$ and $\lambda_k^\Sb(t)$ is associated with certain challenges. More precisely, the expectations $\E{\cdot \mid \Filtration_t^A}$ and $\E{\cdot \mid \Filtration_t^B}$ are taken with respect to the conditional distributions $\pi^A(b, t) = P(\Sb_t=b \mid \Filtration_t^A)$ and $\pi^B(a, t) = P(\Sa_t=a \mid \Filtration_t^B)$, for which analytic expressions are not available. 
We have shown previously \cite{Duso2018} that a stochastic differential equation can be found for such conditional distributions.  
For instance, the equation for $\pi^B(a, t)$ would read 
\begin{equation}
\begin{split}
    \d\pi^B(a, t)&=\sum_{k\in\R{A}} \Big[ h_k(a-\nu_k^{\Sa},B_t)\pi^B(a-\nu_k^{\Sa},t) \\ 
    &\quad\quad\quad\quad\quad -h_k(a,B_t)\pi^B(a,t)\Big] \d t  \\
    &\quad-\sum_{k\in\R{B|A}} \pqty{h_k(a,B_t)-\lambda_k^A(t)}\pi^B(a, t)\d t  \\
    &\quad+\sum_{k\in\R{B|A}} \frac{h_k(a,B_t)-\lambda_k^A(t)}{\lambda_k^A(t)}\pi^B(a,t)\mathrm{d}N_k(t)
\end{split}
\label{eq:Filtering}
\end{equation}
with $\R{\Sb|\Sa}\subset\R{\Sb}$ denoting the set of coupling reactions affecting species $\species{\Sb}$ which are driven by $\species{\Sa}$. In (\ref{eq:Filtering}), the terms $\mathrm{d}N_k(t)$ correspond to the differential version of the reaction counters $N_k$ at time $t$. 
A corresponding equation for $\pi^A(b, t)$ can be formulated analogously. 
Since the propensities obey the mass-action law from (\ref{eq:MassAction}), we further realize that $\lambda_k^\Sa(t)$ and $\lambda_k^\Sb(t)$ are just functions of the moments of the associated filtering distributions. The calculation of the marginal propensities is therefore equivalent to the calculation of the moments of $\pi^A(b, t)$ and $\pi^B(a, t)$. However, depending on the stoichiometry of the considered system, the obtained moment dynamics may not be closed. In this case we can employ suitable moment-closure schemes \cite{hespanha2008moment} to derive approximate equations for the conditional moments.
We will provide explicit expressions for the time-evolution of $\lambda_k^\Sa(t)$ and $\lambda_k^\Sb(t)$ for some examples in Section \ref{sec:casestudies}. For further information on the marginal process framework and the calculation of the conditional expectations the reader may refer to \cite{Duso2018}.

\subsection{Radon-Nikodym derivatives and Jacod's formula}
\label{sec:RN}
Having discussed the joint and marginal dynamics of $\Sa_t$ and $\Sb_t$, we are now ready to study the Radon-Nikodym derivative that appears inside~(\ref{eq:PMI}). To this end, we employ Jacod's formula, which provides an explicit form of the Radon-Nikodym derivative for multivariate counting processes. Consider a counting process $\SS_t$ of the form (\ref{eq:RTC}) with natural filtration $\Filtration_t^X$. We define path measures $\MeasureQ^X$ and $\widehat{\MeasureQ}^X$ restricted to $\Filtration_t^X$ under which $\SS$ has propensitities $\eta_k(t)$ and $\hat{\eta}_k(t)$, respectively. Note that $\eta_k(t)$ and $\hat{\eta}_k(t)$ may be history-dependent in general. The Radon-Nikodym derivative of $\MeasureQ^X$ with respect to $\widehat{\MeasureQ}^X$ is given by \cite{Liptser2001}
\begin{equation}
    \frac{\d\MeasureQ^X}{\d\widehat{\MeasureQ}^X} = \frac{\prod_{k=1}^K \prod_{j=1}^{N_k(t)} \eta_k(T_{k,j}^-) e^{-\int_0^t \eta_k(s)\d s} }{\prod_{k=1}^K \prod_{j=1}^{N_k(t)} \hat{\eta}_k(T_{k,j}^-) e^{-\int_0^t \hat{\eta}_k(s) \d s} },
\end{equation}
where the symbol $T_{k,j}^-$ is the left limit to the $j^{th}$ firing time of reaction channel $k$.
Therefore, if we set $\MeasureQ^X$ and $\widehat{\MeasureQ}^X$ to $\Measure^{AB}$ and $\Measure^A \times \Measure^B$, respectively, we obtain for the Radon-Nikodym derivative inside (\ref{eq:PMI})
\begin{equation}
\begin{split}
    \frac{\d\Measure^{AB}}{\d(\Measure^{A}\times \Measure^{B})} &= \prod_{k=1}^K \prod_{j=1}^{N_k(t)} \lambda_k^{AB}(T_{k,j}^-) e^{-\int_0^t \lambda_k^{AB}(s)\d s}  \\
    &\quad \times  \Bigg( \prod_{k\in \R{\Sa}} \prod_{j=1}^{N_k(t)} \lambda_k^{A}(T_{k,j}^-) e^{-\int_0^t \lambda_k^{A}(s)\d s}  \\
    &\quad \times \prod_{k\in \R{\Sb}} \prod_{j=1}^{N_k(t)} \lambda_k^{B}(T_{k,j}^-) e^{-\int_0^t \lambda_k^{B}(s)\d s}\Bigg)^{-1} .
    \end{split}
    \label{eq:RNC}
\end{equation}

\subsection{Path mutual information}
\label{sec:MainResult}
We recognize that the path mutual information defined in~(\ref{eq:PMI}) is just the expectation of the logarithm of~(\ref{eq:RNC}), which reads 
\begin{equation}
    \begin{split}
    &\log \frac{\d\Measure^{AB}}{\d(\Measure^{A}\times \Measure^{B})} = \\ &\quad \sum_{k=1}^K \sum_{j=1}^{N_k(t)} \log \lambda_k^{AB}(T_{k,j}^-) -\int_0^t \lambda_k^{AB}(s)\d s  \\
    &\quad \quad-\sum_{k\in \R{\Sa}} \sum_{j=1}^{N_k(t)} \log \lambda_k^{A}(T_{k,j}^-) +\int_0^t \lambda_k^{A}(s)\d s  \\
    &\quad \quad-\sum_{k\in \R{\Sb}} \sum_{j=1}^{N_k(t)} \log \lambda_k^{B}(T_{k,j}^-) +\int_0^t \lambda_k^{B}(s)\d s.
    \end{split}
    \label{eq:LogRNC}
\end{equation}
Note that (\ref{eq:LogRNC}) can be simplified by realizing that the resulting (finite) sums over the reaction occurrences can be written as stochastic integrals with respect to the differential reaction counters $\mathrm{d}N_k(t)$. 
Before we state the final result, we take into account the fact that only the coupling reactions are affected by the marginalization. The propensities of all other reactions will remain unaffected when $\Filtration^{AB}$ is replaced by $\Filtration^{A}$ or $\Filtration^{B}$, respectively. Those reactions will have two contributions in (\ref{eq:LogRNC}) with opposite sign and will thus cancel out. Taking this into account and performing some minor rearrangements yields
\begin{align}
    &\PMI{\Sa}{\Sb}{t}=\E{\log \frac{\d\Measure^{AB}}{\d(\Measure^{A}\times \Measure^{B})}}
     \notag \\
    &=\mathbb{E}\Big[\sum_{k\in\R{\Sa|\Sb}}\int_0^t \pqty{\log\lambda_k^{AB}(s^-) -\log\lambda_k^A(s^-)}\d N_k(s) \notag \\
    &\quad\quad\quad\quad\quad\quad\quad
    - \int_0^t \pqty{ \lambda_k^{AB}(s) - \lambda_k^A(s) } \d s \notag \\
      &+\sum_{k\in\R{\Sb|\Sa}}\int_0^t \pqty{\log\lambda_k^{AB}(s^-) -\log\lambda_k^B(s^-)}\d N_k(s) \notag \\
    &\quad\quad\quad\quad\quad\quad\quad
    - \int_0^t \pqty{ \lambda_k^{AB}(s) - \lambda_k^B(s) } \d s \Big] ,
    \label{eq:estimateMI}
\end{align}
where we recall that the symbol $\R{\Sa|\Sb}$ denotes the coupling reactions affecting species $\species{\Sa}$ which are driven by $\species{\Sb}$, and vice versa for $\R{\Sb|\Sa}$. 
The symmetric structure of~(\ref{eq:estimateMI}) reflects the fact that information can be transferred both from $\R{\Sa}$ to $\R{\Sb}$ as well as $\R{\Sb}$ to $\R{\Sa}$. 
In order to numerically evaluate~(\ref{eq:estimateMI}), we can use Gillespie's stochastic simulation algorithm to simulate paths of the whole network $\R{\SS}$ and for each of them, evaluate the integrals in (\ref{eq:estimateMI}) using the marginal propensities described in Section~\ref{sec:marginal}. Subsequently, the path-MI is obtained by averaging over all sample paths.


\subsection{Connection to previous results}
\label{sec:previous}
While our study focuses on the explicit calculation of the path-MI from (\ref{eq:estimateMI}), we want to show how our results can be related to existing theoretical work in the literature. 
Following \cite{Solo2005,Guo2008}, we realize that the expectation of the Riemann integrals in~(\ref{eq:estimateMI}) is equal to zero. This can be seen by exchanging the order of the expectation and time integration and realizing that $\E{\lambda_k^{AB}(s)-\lambda_k^{A}(s)}=0$ and $\E{\lambda_k^{AB}(s)-\lambda_k^{B}(s)}=0$ by definition of the marginal propensity. Additionally, we can use the Doob-Meyer decomposition theorem and expand the differential reaction counters in a predictable part and a martingale $\d N_k(s)=\lambda_k^{AB}(s)\d s+\d Q_k(s)$, where $\E{\d Q_k(s) \mid \Filtration_s^{AB}}=0$. Thus, (\ref{eq:estimateMI}) can be reformulated as
\begin{align}
    \PMI{\Sa}{\Sb}{t}=&\sum_{k\in\R{C}}\int_0^t  \E{\lambda_k^{AB}(s)\log\lambda_k^{AB}(s)} \d s
     \notag \\
    &-\sum_{k\in\R{\Sa|\Sb}}\int_0^t  \E{\lambda_k^A(s)\log\lambda_k^A(s)} \d s \notag \\
    &-\sum_{k\in\R{\Sb|\Sa}}\int_0^t  \E{\lambda_k^B(s)\log\lambda_k^B(s)} \d s ,
    \label{eq:expectationMI}
\end{align}
where $\R{C}=\R{\Sa|\Sb}\cup\R{\Sb|\Sa}$ denotes the set of coupling reactions.
Note that~(\ref{eq:expectationMI}) resembles the point process mutual information given in~\cite{Solo2005} and analogous results can be found in~\cite{Guo2008,Pasha2012}. Similar considerations hold also for the transfer entropy~\cite{Spinney2016}, whose definition is  related to~(\ref{eq:PMI}). 

\section{Case Studies}
\label{sec:casestudies}
In the following, we demonstrate the provided estimator of the path mutual information using three case studies.
All simulations have been performed using the programming language julia \cite{julia}. The code is publicly available at \url{https://github.com/zechnerlab/PathMI} .

\subsection{Protein expression network}
\begin{figure}[tb]
      \centering
      \includegraphics[scale=0.38]{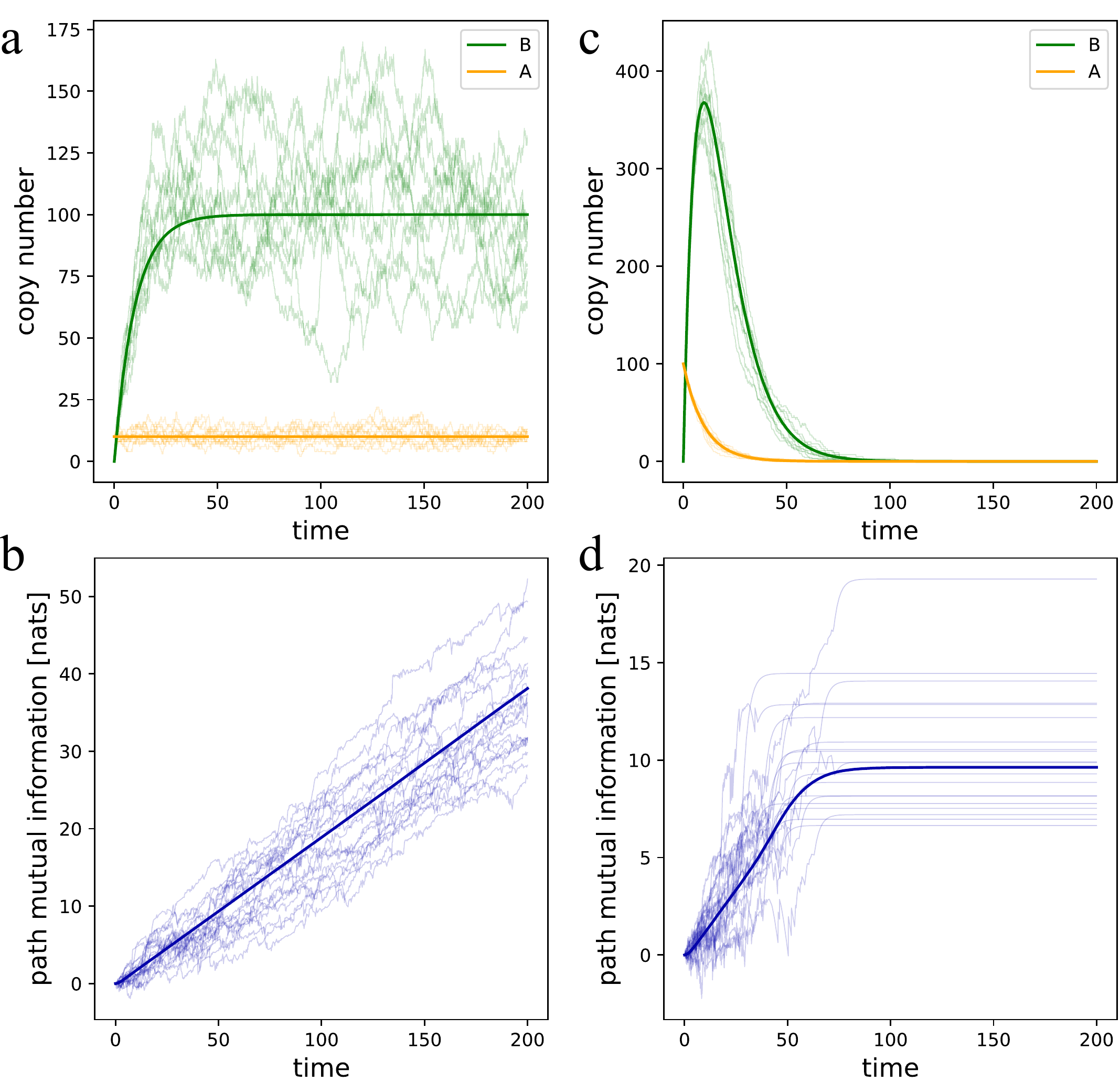}
      \caption{On the left, transcription dynamics (a) and path-MI estimate (b) of the network~(\ref{net:1}) with initial condition $(\Sa_0,\Sb_0)=(10,0)$ and rates $\gamma_\Sa=\gamma_\Sb=1$ and $\delta_\Sa=\delta_\Sb=0.1$. On the right, transient behavior (c) and path-MI estimate (d) of the network~(\ref{net:1}) with initial condition $(\Sa_0,\Sb_0)=(100,0)$ and rates $\gamma_\Sa=0$, $\gamma_\Sb=1$ and $\delta_\Sa=\delta_\Sb=0.1$. In all the plots, thin lines represent single realizations while thick lines are Monte Carlo averages of $10^4$ samples.}
      \label{fig:1}
   \end{figure}

We consider a simple two-stage model of gene expression given by the reaction network
\begin{align}
&\#1:\,\,\,\,\,\,\emptyset \xrightharpoonup{\gamma_\Sa} \species{\Sa} 
&&\#3:\,\,\,\,\,\,\species{\Sa} \xrightharpoonup{\gamma_\Sb} \species{\Sa}+\species{\Sb} \notag \\
&\#2:\,\,\,\,\,\,\species{\Sa} \xrightharpoonup{\delta_\Sa} \emptyset 
&&\#4:\,\,\,\,\,\,\species{\Sb} \xrightharpoonup{\delta_\Sb} \emptyset ,
\label{net:1}
\end{align}
with $\species{A}$ as mRNA, $\species{B}$ as protein and $\gamma_A$, $\gamma_B$, $\delta_A$ and $\delta_B$ the rate constants associated to mRNA and protein synthesis and degradation.
We want to quantify the path mutual information between the mRNA and protein paths. 
In this example the evolution of $\species{\Sa}$ is not driven by $\species{\Sb}$, thus $\R{\Sa|\Sb}=\emptyset$. Therefore, in order to estimate $\PMI{\Sa}{\Sb}{t}$, we only need to focus on reaction $\#3$ and be able to compute the marginal propensity $\lambda_3^B(t)=\gamma_\Sb\Mom{\Sa}{1}(t)$, where $\Mom{\Sa}{1}(t)=\E{A_t|\Filtration_t^B}$ denotes the first moment of the conditional distribution $\pi^B(a,t)$.
Thus, (\ref{eq:estimateMI}) just corresponds to
\begin{align}
    &\PMI{\Sa}{\Sb}{t}=\mathbb{E}\Big[\int_0^t \pqty{\log\pqty{\gamma_\Sb\Sa_s} -\log(\gamma_\Sb\Mom{\Sa}{1}(s)) }\d N_3(s) \notag \\
    &\,\,\,\,\,\,\,\,\,\,\,\,\,\,\,\,\,\,\,\,\,\,\,\,\,\,\,\,\,\,\,\,\,\,\,\,\,\,\,\,
    - \gamma_\Sb\int_0^t \pqty{\Sa_s - \Mom{\Sa}{1}(s)} \d s \Big] ,
    \label{eq:MI_1}
\end{align}
where $\d N_3(t)$ is the differential reaction counter of the coupling reaction. 
In order to evaluate $\Mom{\Sa}{1}$, we instantiate the filtering equation from (\ref{eq:Filtering}) for this particular example, i.e.,
\begin{align}
    \d \pi^{\Sb}(\sa,t)&=\gamma_\Sa\bqty{\pi^{\Sb}(\sa-1,t)-\pi^{\Sb}(\sa,t)}\d t \notag \\
    &+ \delta_\Sa\bqty{(\sa+1)\pi^{\Sb}(\sa+1,t)-\sa\pi^{\Sb}(\sa,t)}\d t \notag \\
    &-\gamma_\Sb\bqty{\sa-\Mom{\Sa}{1}(t)}\pi^{\Sb}(\sa,t)\d t \notag \\
    &+\frac{\sa-\Mom{\Sa}{1}(t)}{\Mom{\Sa}{1}(t)}\pi^{\Sb}(\sa,t)\d N_3(t) .
    \label{eq:filtering1}
\end{align}
We can now obtain a stochastic differential equation for $\Mom{\Sa}{1}(t)=\sum_0^\infty a\,\pi^{\Sb}(\sa,t)$
by multiplying both sides of $(\ref{eq:filtering1})$ by $a$ and summing over all possible values of $a$. This yields
\begin{align}
    \d \Mom{\Sa}{1}=&\bqty{\gamma_\Sa-\delta_\Sa\Mom{\Sa}{1}(t)-\gamma_\Sb\pqty{\Mom{\Sa}{2}(t)-\Mom{\Sa}{1}^2(t)}} \d t \notag \\
    &+\frac{\Mom{\Sa}{2}(t)-\Mom{\Sa}{1}^2(t)}{\Mom{\Sa}{1}(t)}\d N_3(t) .
    \label{eq:ex1_mRNA1}
\end{align}
However, since (\ref{eq:ex1_mRNA1}) depends on higher order moments, we adopt a Gamma closure at second order so that we can write the closed evolution of $\Mom{\Sa}{2}=\E{A_t^2|\Filtration_t^B}$ as
\begin{align}
    \d \Mom{\Sa}{2}=&\Big[\gamma_\Sa+(2\gamma_\Sa+\delta_\Sa)\Mom{\Sa}{1}(t)-\delta_\Sa\Mom{\Sa}{2}(t) \notag \\ &-2\gamma_\Sb\frac{\Mom{\Sa}{2}(t)}{\Mom{\Sa}{1}(t)}\pqty{\Mom{\Sa}{2}(t)-\Mom{\Sa}{1}^2(t)}\Big] \d t \notag \\
    &+2\frac{\Mom{\Sa}{2}(t)}{\Mom{\Sa}{1}^2(t)}\pqty{\Mom{\Sa}{2}(t)-\Mom{\Sa}{1}^2(t)}\d N_3(t) .
    \label{eq:ex1_mRNA2}
\end{align}
Equations~(\ref{eq:ex1_mRNA1}) and~(\ref{eq:ex1_mRNA2}) can be solved alongside of the SSA simulations in order to evaluate~(\ref{eq:MI_1}). Finally, $\PMI{\Sa}{\Sb}{t}$ can be estimated as a Monte Carlo over multiple SSA simulations.
Fig.~\ref{fig:1}(a-b) shows simulation results for this example. The mRNA and protein levels were initialized to $\gamma_A/\delta_A$ and zero respectively. After a short initial transient, the path mutual information increases linearly with time, indicating that information is transferred from $\mathrm{A}$ to $\mathrm{B}$ at constant rate as soon as the system approaches its stationary state. Individual realizations are shown to illustrate their variability around the linear trend.

\subsection{Transient induction of transcription}
We again consider the reaction network introduced in~(\ref{net:1}) but in a different regime. In particular we assume that transcription is switched off at time zero such that $\gamma_\Sa=0$. The initial pool of mRNA $\Sa_0>0$ will then degrade as time increases and the system will ultimately converge to a zero steady state. 
Our goal is to quantify the information transmission between species $\species{\Sa}$ and $\species{\Sb}$ during this transient period. 
Note that the expressions~(\ref{eq:MI_1}), (\ref{eq:filtering1}), (\ref{eq:ex1_mRNA1}) and~(\ref{eq:ex1_mRNA2}) still apply, with the only difference that $\gamma_\Sa=0$. In Fig.~\ref{fig:1}(c-d) we show the results for the initial conditions $\Sa_0=100$ and $\Sb_0=0$. The results show that the path-MI stops to increase after both the species are extinct. First, the mRNA gets depleted and consequently the coupling reaction $\#3$ stops firing. After this point, the $log$-jump contributions in~(\ref{eq:MI_1}) cease and the residual contribution only comes from the integral of the difference between $\Sa_s=0$ and its causal estimate $\Mom{\Sa}{1}(s)$ in the second line of~(\ref{eq:MI_1}). 

\subsection{A stochastic oscillatory system}
In our last case study, we consider a stochastic predator-prey system inspired by the Lotka-Volterra model. The system is defined by the reaction network
\begin{align}
&\#1:\,\,\,\,\,\,\species{\Sa} \xrightharpoonup{\gamma_\Sa} 2\species{\Sa} 
&&\#3:\,\,\,\,\,\,\species{\Sa}+\species{\Sb} \xrightharpoonup{\gamma_\Sb} \species{\Sa}+2\species{\Sb} \notag \\
&\#2:\,\,\,\,\,\,\species{\Sa}+\species{\Sb} \xrightharpoonup{\delta_\Sa} \species{\Sb} 
&&\#4:\,\,\,\,\,\,\species{\Sb} \xrightharpoonup{\delta_\Sb} \emptyset,
\label{net:3}
\end{align}
where the species $\species{\Sa}$ plays the role of the prey and $\species{\Sb}$ is the predator. 
A prey can duplicate with reproduction rate $\gamma_\Sa$ and can be consumed by a predator with rate $\delta_\Sa$. Instead, a predator can die with death rate $\delta_\Sb$ and duplicate proportionally to the amount of prey, with rate $\gamma_\Sb$.

Notice that in this example we have $\R{\Sa}=(\#1,\#2)$ and $\R{\Sb}=(\#3,\#4)$ and two coupling reactions, $\R{\Sa|\Sb}=\#2$ and $\R{\Sb|\Sa}=\#3$. It is worth mentioning that reaction $\#3$ satisfies the smooth coupling assumption because only $\species{\Sb}$ gets modified by it. 
  \begin{figure}[tb]
      \centering
      \includegraphics[scale=0.47]{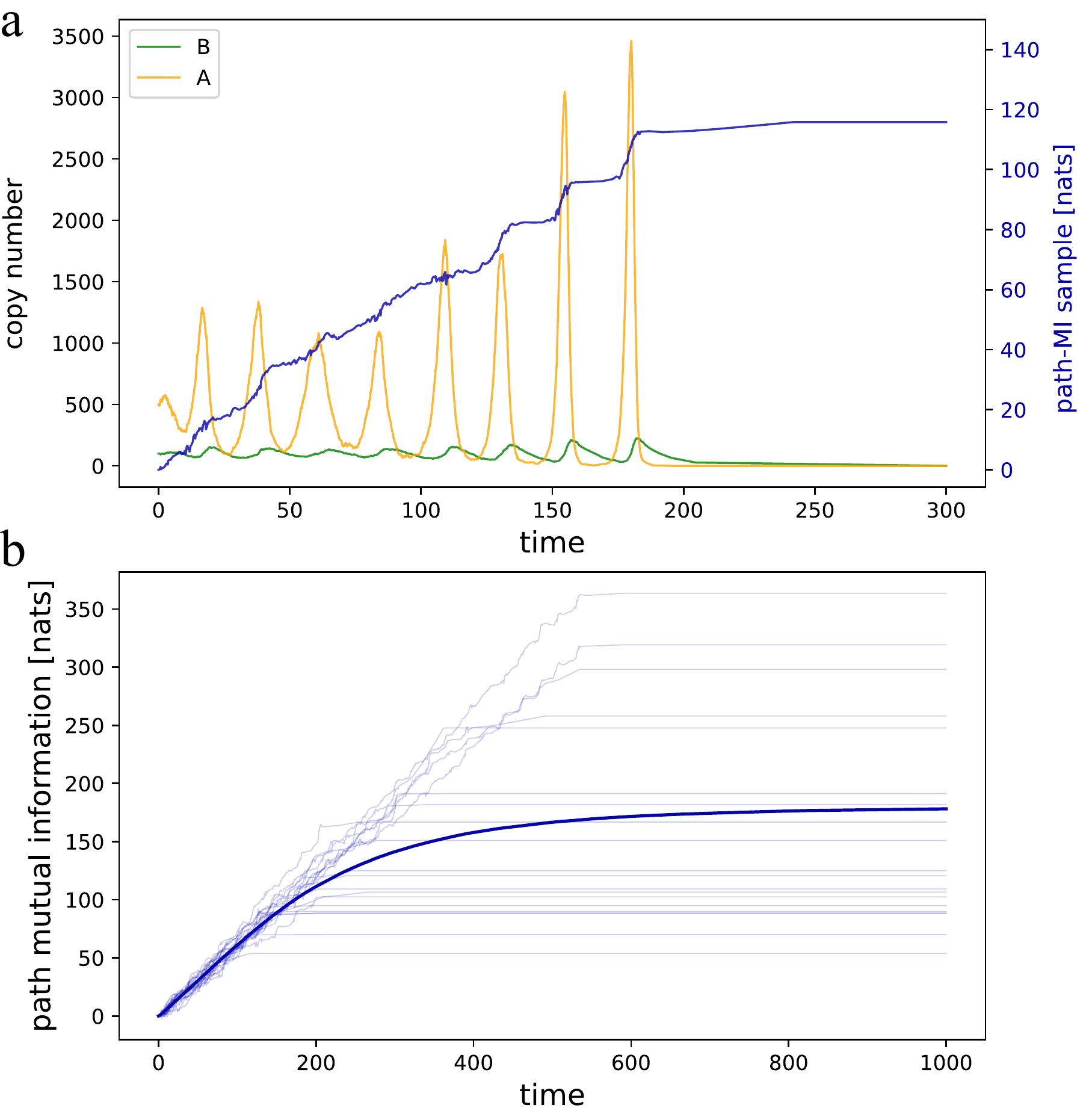}
      \caption{Behavior of the network~(\ref{net:3}) with initial condition $(\Sa_0,\Sb_0)=(500,100)$ and rates $\gamma_\Sa=1$, $\delta_\Sa=0.01$, $\gamma_\Sb=2\cdot 10^{-4}$ and $\delta_\Sb=0.1$. (a) A single realization of the system and the corresponding path-MI sample. (b) Path mutual information (thick line) estimated using $10^3$ samples. Several individual samples are shown as thin lines.}
      \label{fig:2}
   \end{figure}
Note that, since in~(\ref{net:3}) the information transfer happens bi-directionally between $\Sa$ and $\Sb$, we require expressions for both the marginal propensities $\lambda_2^A(t)$ and $\lambda_3^B(t)$. Due to space considerations, we omit explicit expressions for $\pi^\Sa$ and $\pi^\Sb$ and the conditional moments, which have been obtained again under a Gamma closure. In Fig.~\ref{fig:2}(a) we show a realization of the system and the corresponding path-MI sample. The trajectories of prey and predator exhibit a characteristic oscillatory pattern. The path-MI sample grows significantly during the copy number peaks of prey and predator, because of the increased intensity of the coupling reactions. In this model it might happen that the preys go extinct in which case information transmission stops. 
As time increases, more and more realizations reach extinction, which causes the path-MI to saturate (Fig.~\ref{fig:2}(b)). 

\addtolength{\textheight}{-3cm}   

\section{Conclusions}

In this work we have presented a method to efficiently estimate the path mutual information between two chemical species whose dynamics evolve according to a stochastic reaction network. In order to derive the scheme, we employed our recently proposed marginal process framework, which allowed us to explicitly calculate the required Radon-Nikodym derivatives. We showed how the path-MI can be efficiently estimated by combining SSA with the solution of a stochastic filtering problem. 
We showed the efficacy of the method by applying it to three case studies with different dynamical properties. 
For the purposes of this work we restricted ourselves to a specific class of reaction systems, which consist of only two smoothly coupled species. In the future, we will extend the method to networks comprising more than two chemical species and arbitrary coupling reactions.

\section{Acknowledgments}
The authors were supported by core funding of the Max Planck Institute of Molecular Cell Biology and Genetics and the Center for Systems Biology Dresden and the Center for Advancing Electronics Dresden (cfaed).

\footnotesize               
\bibliographystyle{ieeetr}  

\end{document}